\documentclass[reprint,
superscriptaddress,
 amsmath,amssymb,
 aps,
floatfix,
longbibliography
]{revtex4-1}
\usepackage{amsmath}
\usepackage{graphicx}
\usepackage{multirow}
\usepackage{dcolumn}
\usepackage{bm}

\begin{document}

\title{Nanomechanical probing and strain tuning of the Curie temperature in suspended Cr$_2$Ge$_2$Te$_6$ heterostructures}

\author{Makars \v{S}i\v{s}kins}
\author{Samer Kurdi}
\author{Martin Lee}
\author{Benjamin J. M. Slotboom}
\affiliation{%
 Kavli Institute of Nanoscience, Delft University of Technology, Lorentzweg 1,\\
 2628 CJ, Delft, The Netherlands
}

\author{Wenyu Xing}
\affiliation{%
International Center for Quantum Materials, Peking University, 5 Summer Palace Road, \\
100871 Beijing, China
}

\author{Samuel Ma\~{n}as-Valero}
\author{Eugenio Coronado}
\affiliation{%
Instituto de Ciencia Molecular (ICMol), Universitat de Val\`{e}ncia, c/Catedr\'{a}tico Jos\'{e} Beltr\'{a}n 2,\\ 46980 Paterna, Spain
}%
\author{Shuang Jia}
\author{Wei Han}
\affiliation{%
International Center for Quantum Materials, Peking University, 5 Summer Palace Road, \\
100871 Beijing, China
}

\author{Toeno van der Sar}
\author{Herre S. J. van der Zant}
\affiliation{%
 Kavli Institute of Nanoscience, Delft University of Technology, Lorentzweg 1,\\
 2628 CJ, Delft, The Netherlands
}

\author{Peter G. Steeneken}
\email{e-mail: m.siskins-1@tudelft.nl; t.vandersar@tudelft.nl; h.s.j.vanderzant@tudelft.nl; p.g.steeneken@tudelft.nl}
\affiliation{%
	Kavli Institute of Nanoscience, Delft University of Technology, Lorentzweg 1,\\
	2628 CJ, Delft, The Netherlands
}%
\affiliation{
	Department of Precision and Microsystems Engineering, Delft University of Technology,
	Mekelweg 2,\\ 2628 CD, Delft, The Netherlands}

%

\begin{abstract}
Two-dimensional (2D) magnetic materials with strong magnetostriction are interesting systems for strain-tuning the magnetization, enabling potential for realizing spintronic and nanomagnetic devices. Realizing this potential requires understanding of the magneto-mechanical coupling in the 2D limit. In this work, we suspend thin Cr$_2$Ge$_2$Te$_6$ layers, creating nanomechanical membrane resonators. We probe its mechanical and magnetic properties as a function of temperature and strain. Pronounced signatures of magneto-elastic coupling are observed in the temperature-dependent resonance frequency of these membranes near $T_{\rm C}$. We further utilize Cr$_2$Ge$_2$Te$_6$ in heterostructures with thin layers of WSe$_2$ and FePS$_3$, which have positive thermal expansion coefficients, to compensate the negative thermal expansion coefficient of Cr$_2$Ge$_2$Te$_6$ and quantitatively probe the corresponding $T_{\rm C}$. Finally, we induce a strain of $0.016\%$ in a suspended heterostructure via electrostatic force and demonstrate a resulting enhancement of $T_{\rm C}$ by $2.5 \pm 0.6$ K in the absence of an external magnetic field. 
\end{abstract}
\maketitle

The recent discovery of long range order in two-dimensional (2D) (anti)ferromagnets \cite{CrI3Huang2017,DiscoveryGong2017,FePS3Lee2016} has triggered extensive studies of 2D materials to experimentally probe magnetism of reduced dimensionality \cite{Gibertini2019}. One material of particular interest is Cr$_2$Ge$_2$Te$_6$ (CGT) - a semiconducting ferromagnet with a bulk Curie temperature, $T_{\rm C}$, $\sim60-66$ K \cite{DiscoveryGong2017,Carteaux1995} with inter- and intra-layer ferromagnetic coupling for any number of layers \cite{DiscoveryGong2017}. Recent progress has been made in manipulating the magnetic order of CGT using electrostatic gating \cite{CGTGateAnisotropyVerzhbitskiy2020,EFieldWang2018}, magnetic field \cite{DiscoveryGong2017, HFieldSelter2020}, pressure \cite{PressureSun2018, PressureSakurai2021}, ion intercalation \cite{IonsWang2019}, and via spin-orbit torque \cite{SpinOrbitGupta2020,SpinOrbitOstwal2020}. Mechanical strain offers another degree of freedom for such manipulation as bulk CGT was recently shown to exhibit strong spin-lattice coupling \cite{PressureSun2018,MagnetostrictionTian2016} and a negative thermal expansion coefficient near $T_{\rm{C}}$ \cite{Carteaux1995}, which is common amongst bulk chromium-based magnetic van der Waals crystals, like CrI$_3$, CrCl$_3$, CrBr$_3$ and Cr$_2$Si$_2$Te$_6$ \cite{Kozlenko2021, Casto2015, McGuire2015,CrCl3NTESchneeloch2021}. However, the coupling between magnetic order and strain in thin CGT has not been studied experimentally, as strain is a difficult parameter to control in substrate-supported ultrathin layers \cite{StraintronicsMiao2021,StrainYang2021}.
 
Emerging nanomechanical methods allow for high-precision strain manipulation and control when 2D materials are suspended forming ultrathin membrane resonators \cite{GrapheneResonatorChen2009,DynamicStrainZhang2020}. Due to the combination of low mass with high strength, these membranes find potential use in high-performance devices \cite{GrapheneResonatorChen2009,GrapheneTunableOscillatorChen2013} and in a wide range of sensor applications \cite{SensorsLemme2020}. The resonance frequency of these membranes can be tuned over a large range by strain \cite{GrapheneResonatorChen2009}, which can be controlled both statically \cite{Lee2008} and dynamically \cite{GrapheneResonatorChen2009,DynamicStrainZhang2020}. Moreover, the difference between the thermal expansion coefficient of suspended 2D material layers and the substrate \cite{Siskins2020,Morell2016} or other materials in a heterostructure \cite{HeteroFengYe2017,kinFaiMakJiang2020} provides additional routes for strain engineering, leading to either stretching or compressing ultrathin layers. Nanomechanical strain can also be used to probe magnetic states of membrane material \cite{Siskins2020} and switch between these states \cite{kinFaiMakJiang2020, NeelVectorNi2021}. It is the relation between strain, a material’s thermal expansion coefficient and its magnetic properties which results in the coupling between mechanical and magnetic degrees of freedom, that allows investigation of magnetic phase transitions in 2D layered material membranes \cite{Siskins2020,kinFaiMakJiang2020}. 
 
\begin{figure}
  \begin{center}
  \includegraphics[width=\linewidth]{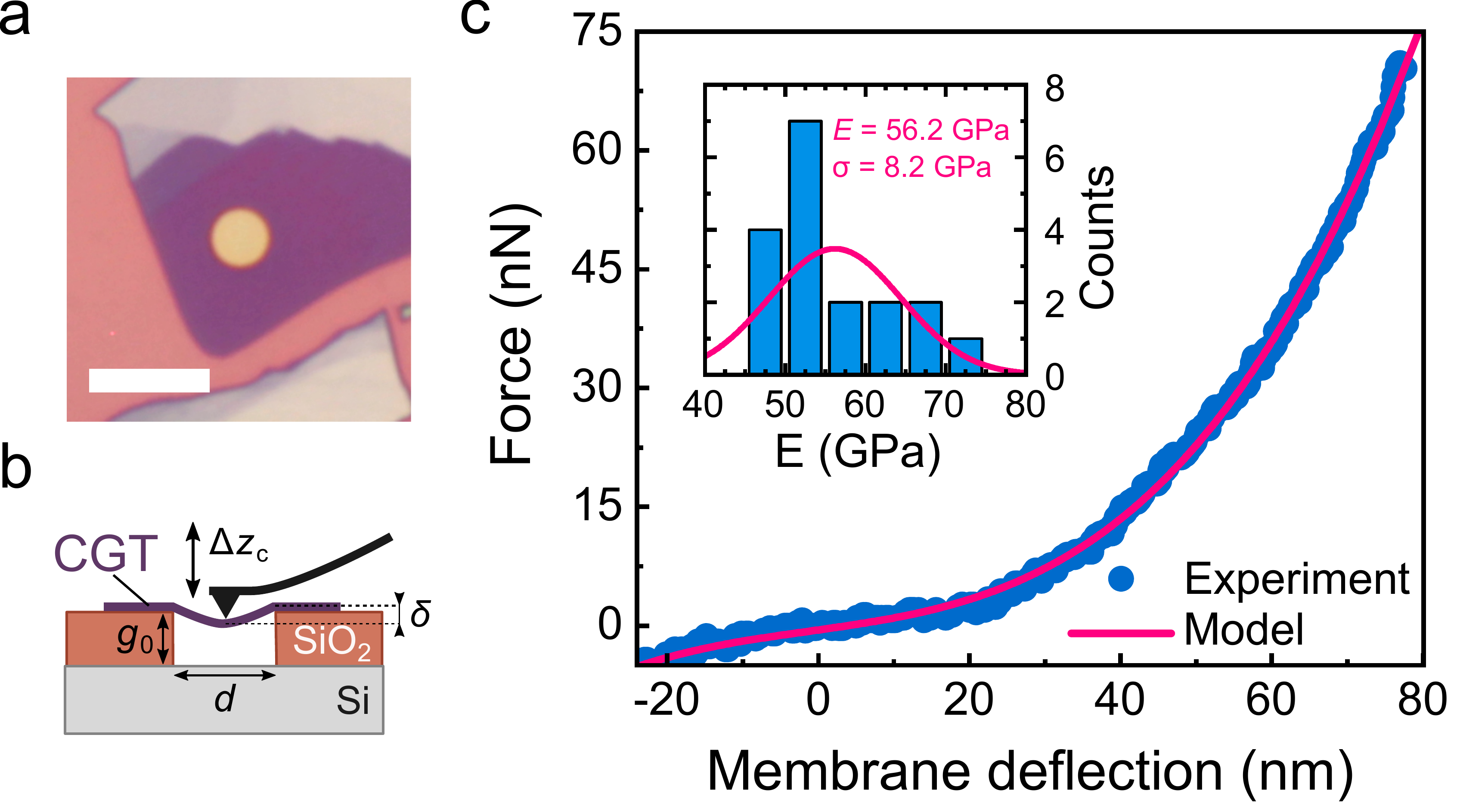}
  \caption{Force indentation of suspended membranes of CGT. \textbf{a} An optical image of a $9.5\pm0.3$ nm thin CGT membrane with a radius $r=2$ $\mu$m. Scale bar: $8$ $\mu$m. \textbf{b} A schematic of the membrane deflection by a cantilever tip. $\Delta z_{\rm c}$ is the cantilever deflection, $g_0=285$ nm the separation between the membrane and the bottom Si substrate, $d$ the membrane diameter, and $\delta$ the membrane deflection. \textbf{c} A force versus membrane deflection plot. Experimental data (filled blue circles) are fit by the point-force deflection of a circular membrane model of equation~(\ref{eq_force_d}) (solid magenta line). The inset shows a histogram of the Young’s moduli extracted from $18$ CGT membranes of $9.5$ - $30.1$ nm thicknesses, which were used to obtain the mean value of the Young’s modulus $E$ and corresponding standard deviation $\sigma$.}
  \label{fig:first}
  \end{center}
\end{figure}

Here, we demonstrate that the magnetic phase transition of suspended thin CGT membranes can be probed mechanically via the temperature-dependent resonance frequency. CGT is known to exhibit high magnetostriction, i.e. a strong magneto-elastic coupling between magnetic order and the lattice \cite{Carteaux1995,MagnetostrictionTian2016,PressureSun2018}, that deforms the crystal near $T_{\rm C}$ \cite{Carteaux1995}. Thus, to control the strain in thin CGT flakes and deterministically probe the transition temperature, we choose other 2D material layers, WSe$_2$ and FePS$_3$, with positive thermal expansion coefficient and integrate them to form heterostructure membranes with CGT \cite{HeteroFengYe2017,HeteroLiu2014,HeteroKim2018,HeterostructureNovoselov2016}, which are important to compensate CGT's negative thermal expansion \cite{Carteaux1995}. We then use these heterostructure membranes to probe and analyze the effect of the mutual interaction between the different 2D materials and CGT, and study the effect of electrostatically induced strain on the ferromagnetic order near $T_{\rm{C}}$.

\section*{Results}
\subsection*{Mechanical properties of CGT membranes.} The force-deflection curve of suspended membranes contains information about mechanical properties of the material. Thus, we first study thin layers of CGT by a static deflection method \cite{Lee2008, CastellanosGomez2012}. We fabricate a freestanding membrane by suspending a mechanically exfoliated thin CGT flake over a circular hole (Fig.~\ref{fig:first}a). We then apply the atomic force microscopy (AFM) force nanoindentation method \cite{Lee2008,CastellanosGomez2012} to indent the centre of the membrane with the tip of an AFM cantilever to cause a deflection of the membrane $\delta$, as shown in Fig.~\ref{fig:first}b. The force applied to the centre of the membrane, $F$, is proportional to the stiffness $k_{\rm c}$ of the cantilever used and its deflection, $\Delta z_{\rm c}$. Using cantilevers with calibrated stiffnesses, we record the force versus membrane deflection curves of this membrane, as depicted in Fig.~\ref{fig:first}c with filled blue circles. The observed trend can be described by the point-force deflection model for a circular membrane, assuming negligible tip radius compared to the membrane diameter \cite{Lee2008,CastellanosGomez2012}:
\begin{equation}
    F=\left(\frac{4\pi E}{3(1-\nu^2)}\frac{h^3}{r^2}\right)\delta+\left(n_0\pi \right)\delta+\left(\frac{q^3Eh}{r^2}\right)\delta^3,
\label{eq_force_d}
\end{equation}
where $E$ is the Young's modulus of the membrane, $r$ the membrane radius, $h$ the membrane thickness, $\nu=0.22$ the Poisson ratio of CGT \cite{deJong2015}, $n_0$ the pre-tension and $q=1/(1.05-0.15\nu-0.16\nu^2)$ is a dimensionless geometrical parameter. To our knowledge, the Young's modulus of CGT has not been experimentally studied. We therefore extract its Young's modulus $E_{\rm{CGT}}$ from the experimental data using equation~(\ref{eq_force_d}), as shown in Fig.~\ref{fig:first}c by the solid magenta line. To obtain a more reliable estimate of $E_{\rm{CGT}}$, we measure $18$ different membranes of varying thicknesses ($h=9.5$ - $30.1$ nm) and radii ($r=1$ - $2.5$ $\mu$m). We find a mean value, $E_{\rm{CGT}}=56.2\pm8.2$ GPa, as shown in the inset of Fig.~\ref{fig:first}c. The obtained value of $E_{\rm{CGT}}$ is consistent with the calculated \cite{Li2014} 2D Young's modulus of CGT, $E_{\rm{2D}}=41.8$ Nm$^{-1}$, which yields to $E=E_{\rm{2D}}/h_{\rm{2D}}=61.0$ GPa, in which $h_{\rm{2D}}$ is the single-layer thickness, taken as a third of that of the unit cell of CGT \cite{Carteaux1995}: $h_{\rm{2D}}=2.056$ nm$/3\approx0.685$ nm. The standard deviation for our measurements is comparable to that reported in similar experiments on other 2D material membranes \cite{Lee2008,CastellanosGomez2012}. We also found no layer number dependence for the studied range of thicknesses \cite{CrI3CantosPrieto2021}.

\begin{figure*}
\begin{center}
  \includegraphics[width=\linewidth]{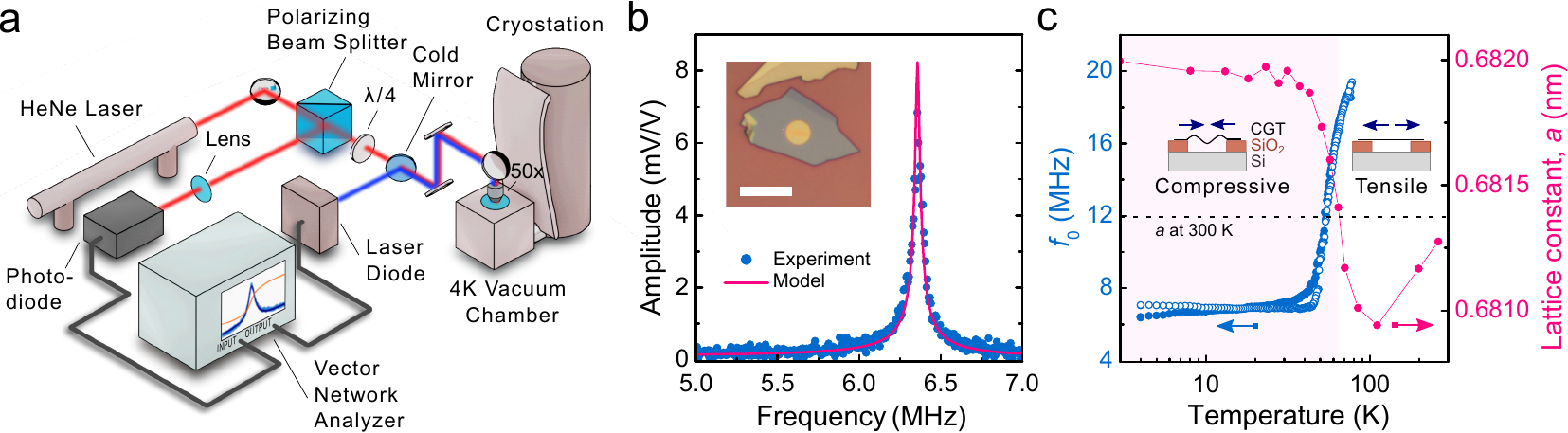}
  \caption{CGT membrane characterization using laser interferometry. \textbf{a} Schematic diagram of the laser interferometry setup. \textbf{b} Filled blue circles - measured amplitude of the fundamental resonance peak at $T=4$ K. Solid magenta line - fit to a linear harmonic oscillator model. The inset shows an optical image of a $17.4\pm0.3$ nm thin CGT membrane of $2$ $\mu$m radius. Scale bar: $8$ $\mu$m. \textbf{c} The comparison of the measured resonance frequency $f_0$ of the thin film shown in \textbf(b) and the lattice constant of bulk CGT as a function of temperature. The filled and open blue circles represent the measured data during heating and cooling cycles respectively. The connected magenta circles represent the lattice constant $a$ of bulk CGT as measured by Carteaux et al \cite{Carteaux1995}. The dashed black horizontal line represents $a$ at $T=300$ K. The insets show schematics of membrane stretching or buckling at tensile or compressive strain, on the left and right respectively. The pink region indicates the temperature range where compressive strain is dominant and is a guide to an eye.}
  \label{fig:second}
  \end{center}
\end{figure*}

\subsection*{Resonating membranes.} In order to study the coupling between the magnetic phase and the mechanical motion, we further investigate the dynamic nanomechanical properties of these membranes as a function of temperature. The temperature dependence of the mechanical resonances of the magnetic membranes is sensitive to changes in the magnetisation of the 2D layers \cite{kinFaiMakJiang2020} and the magnetic phase, via a mechanism that couples the specific heat to the membrane tension via the thermal expansion coefficient \cite{Siskins2020}. Thus, by observing changes in motion of the membrane at $T_{\rm{C}}$ it is possible to probe the ferromagnetic to paramagnetic transition via the mechanical resonance frequency \cite{Siskins2020}. To do this, we use a laser interferometry technique \cite{Siskins2020, MoS2ResonatorCastellanosGomez2013} (Fig.~\ref{fig:second}a and Experimental Section). A CGT membrane (see the inset of Fig.~\ref{fig:second}b) is placed in the chamber of an optical cryostation. We use a power-modulated blue laser to opto-thermally excite a fundamental resonance mode of the membrane, and a red laser to measure the change in the reflected signal due to the subsequent displacement of the membrane. 

We first measure the resonance peak of the fundamental membrane mode at $T=4$ K (blue circles in Fig.~\ref{fig:second}b). We fit the experimental data to a harmonic oscillator model (solid magenta line in Fig.~\ref{fig:second}b) to determine the frequency of the fundamental membrane mode, $f_0$. Subsequently, while recording $f_0(T)$ we heat the sample to $T=78$ K, above the expected $T_{\rm C}$ of $66$ K \cite{DiscoveryGong2017,Carteaux1995}, and cool it down to $T=4$ K. We plot the experimental data for the heating (filled blue circles) and the cooling (open blue circles) cycle in Fig.~\ref{fig:second}c. As the CGT membrane is cooled through the $T_{\rm C}$, its resonance frequency $f_0$ reduces from $19.3$ MHz at $78$ K to $6.3$ MHz at $4$ K. To describe the $f_0(T)$ behaviour of CGT, we model the resonance frequency of a circular membrane as:
\begin{equation}
    f_0(T)=\sqrt{\left(\frac{2.4048}{2\pi r}\right)^2\frac{n_{\rm{th}}(T)}{\rho h}+f_0^2(T_0)},
    \label{eq:frequency}
\end{equation}
where $f_0(T_0)$ is the resonance frequency at a reference temperature $T_0$ (e.g. room temperature) due to the contribution of the pre-tension and the bending rigidity, $\rho$ the mass density, $n_{\rm{th}}(T)=\frac{Eh}{(1-\nu)}\epsilon_{\rm{th}}$ the thermally accumulated tension, $\epsilon_{\rm{th}}=-\int_{T_0}^{T_i}\left( \alpha_{\rm{CGT}}(T)-\alpha_{\rm{Si}}(T)\right) \, \mathrm{d}T$ the thermal strain at an arbitrary temperature $T_{\rm i}$ \cite{Siskins2020,Morell2016}, $\alpha_{\rm Si}(T)$ the literature values for thermal expansion coefficient of Si substrate \cite{Lyon1977}, and $\alpha_{\rm{CGT}}(T)$ the temperature dependent thermal expansion coefficient of CGT. Thus, we attribute the observed $f_0(T)$ trend to a large change in the in-plane lattice constants of the unit cell and the resulting negative $\alpha_{\rm CGT}(T)$ near the phase transition, which was also previously reported for bulk CGT \cite{Carteaux1995}. This anomalous lattice expansion, when cooling down from the paramagnetic to the ferromagnetic state, is related to the strong magnetostriction effect at the ferromagnetic ordering temperature $T_{\rm C}$ in CGT \cite{Carteaux1995, MagnetostrictionTian2016} that causes a substantial drop in strain $\epsilon_{\rm{th}}(T)$ and a corresponding reduction of the membrane tension near $T_{\rm C}$. 

In Fig.~\ref{fig:second}c we also plot the change of the in-plane lattice constant, $a$, of bulk CGT, experimentally obtained by Carteaux et al. \cite{Carteaux1995}, as a function of temperature and compare it with the observed resonance frequency. The lattice constant $a$ is shown to be smaller at room temperature (indicated by the black horizontal dashed line) in comparison to temperatures below the paramagnetic to ferromagnetic phase transition ($\sim66$ K). Since the lattice constant increases, the suspended part of the flake effectively has a larger volume at $T<T_{\rm C}$ than at room temperature, indicating a switch from tensile to compressive strain, as indicated by the pink region in Fig.~\ref{fig:second}c. In this temperature range, the compressively strained suspended flake is likely to sag, irreproducibly wrinkle or buckle, which possibly explains a different warming and cooling trend for the resonance frequency $f_0(T)$. Buckling of magnetic layers itself is interesting for the development of reprogrammable mechanical memory devices at nano- and micro-scale \cite{MechanicalMemoryChen2021} utilizing the bi-stable states of a buckled flake \cite{BistableMahboob2008,BistableRoodenburg2009}. However, due to the changing resonance mode shape in the compressive strain regime, and the potential wrinkling, equation~(\ref{eq:frequency}) is not reliable for $T<T_{\rm C}$. Thus, a comprehensive analysis of $f_0(T)$ near and below the phase transition \cite{Siskins2020} cannot be applied to resonators made of bare CGT. 

\begin{figure*}[ht]
\begin{center}
  \includegraphics[width=28pc]{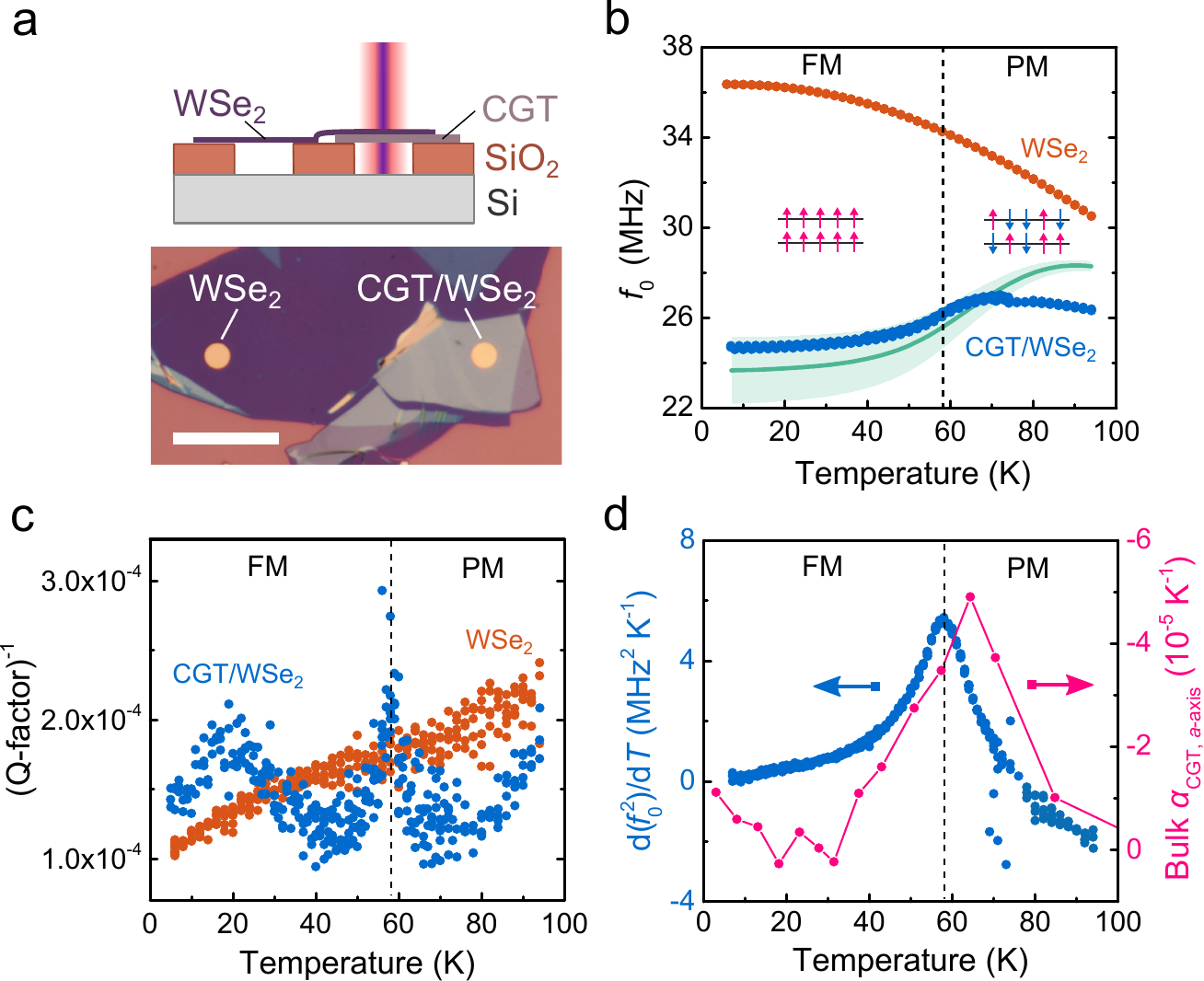}
  \caption{Mechanical properties of a suspended CGT/WSe$_2$ ($16.7\pm0.3$ nm/$6.9\pm0.1$ nm) heterostructure membrane with a radius $r=2$ $\mu$m. \textbf{a} Top panel: Schematic cross-section of the sample consisting of the suspended CGT/WSe$_2$ heterostructure membrane and the reference WSe$_2$ membrane of the same flake. Bottom panel: Optical image of the sample with specific membranes indicated. Scale bar: $16$ $\mu$m. \textbf{b} Measured resonance frequencies $f_0$ of the membranes in (a). Solid green line - the fitted model of the resonance frequency of the CGT/WSe$_2$ heterostructure (equation~(\ref{eq_f_hetero})). Light green region - the allowed higher and lower boundary of the model due to the uncertainties in $h_{1,2}$, $E_{1,2}$ and $f^2_0(T_0)$. Insets: Schematic of the collinear and random magnetic spin arrangement in a bilayer of CGT in the ferromagnetic (FM) and the paramagnetic (PM) phases respectively. \textbf{c} Mechanical damping $Q^{-1}$ as a function temperature. \textbf{d} Filled blue circles - $\frac{\mathrm{d}\left(f_{0}^2\right)}{\mathrm{d}T}$ of the CGT/WSe$_2$ heterostructure as a function of temperature. Connected magenta circles - the thermal expansion coefficient of bulk CGT \cite{Carteaux1995}. The black vertical dashed lines in (b-d) indicate the measured $T^*_{\rm C}=58\pm0.5$ K as determined from the peak position of $\frac{\mathrm{d}\left(f_{0}^2\right)}{\mathrm{d}T}$.}
  \label{fig:third}
\end{center}
\end{figure*}

\subsection*{Tensile strain in heterostructure membranes.} To prevent the compressive strain that can lead to wrinkling in CGT membranes, and controllably probe $f_0(T)$ near and below its phase transition temperature, we fabricate a suspended membrane heterostructure composed of CGT and WSe$_2$ flakes, shown in Fig.~\ref{fig:third}. WSe$_2$ is a material with well-known mechanical properties \cite{Morell2016, WSe2Zhang2016, WSe2Cakir2014} that does not undergo a phase transition from $4$ to $300$ K \cite{WSe2Cakir2014}. In the heterostructure, its positive thermal expansion coefficient $\alpha_{\rm WSe_2}(T)$ counteracts the negative $\alpha_{\rm CGT}(T)$ near $T_{\rm C}$ of CGT, such that the total membrane stress remains tensile and $f_0$ can be used to probe $\alpha(T)$ even below $T_{\rm C}$. 

We measure $f_0(T)$ of the suspended CGT/WSe$_2$ heterostructure together with a reference drum of the same WSe$_2$ flake (Fig.~\ref{fig:third}b). The $f_{0,\rm WSe_2}(T)$ follows a monotonous increase trend with decreasing temperature (filled orange circles), as expected from its positive $\alpha_{\rm WSe_2}(T)$ \cite{Morell2016,WSe2Cakir2014}. In contrast, $f_{0,\rm CGT/WSe_2}(T)$ has a downturn with decreasing temperature near the transition temperature (filled blue circles). This behaviour is distinct from the bare CGT resonator (Fig.~\ref{fig:second}c) and is due to the positive thermal expansion coefficient of the WSe$_2$ layer, which maintains a tensile total strain of the heterostructure. We also plot the mechanical dissipation, the inverse of a quality factor $Q^{-1}(T)$, of both the WSe$_2$ and CGT/WSe$_2$ resonators in Fig.~\ref{fig:third}c. A notable peak in $Q_{\rm CGT/WSe_2}(T)$ is visible at $T=58$ K. We attribute this observation to an increase of the thermoelastic damping \cite{Lifshitz2000, Sun2006} expected near $T_{\rm C}$ in magnetic resonators \cite{Siskins2020}. This observation is supported by the measured maximum in $\frac{\mathrm{d}\left(f_{0}^2\right)}{\mathrm{d}T}$ (filled blue circles) \cite{Siskins2020} at $58\pm0.5$ K, the temperature which we define as $T^*_{\rm C}$ and indicate by the black vertical dashed lines in Fig.~\ref{fig:third}d. The observed maximum at $T^*_{\rm C}$ relates to the peak in the thermal expansion coefficient of CGT at $T_{\rm C}$ (filled magenta circles) that is shifted to lower temperatures in comparison to its bulk values of $60-66$ K \cite{DiscoveryGong2017,Carteaux1995}.

\begin{figure*}
\begin{center}
  \includegraphics[width=0.85\linewidth]{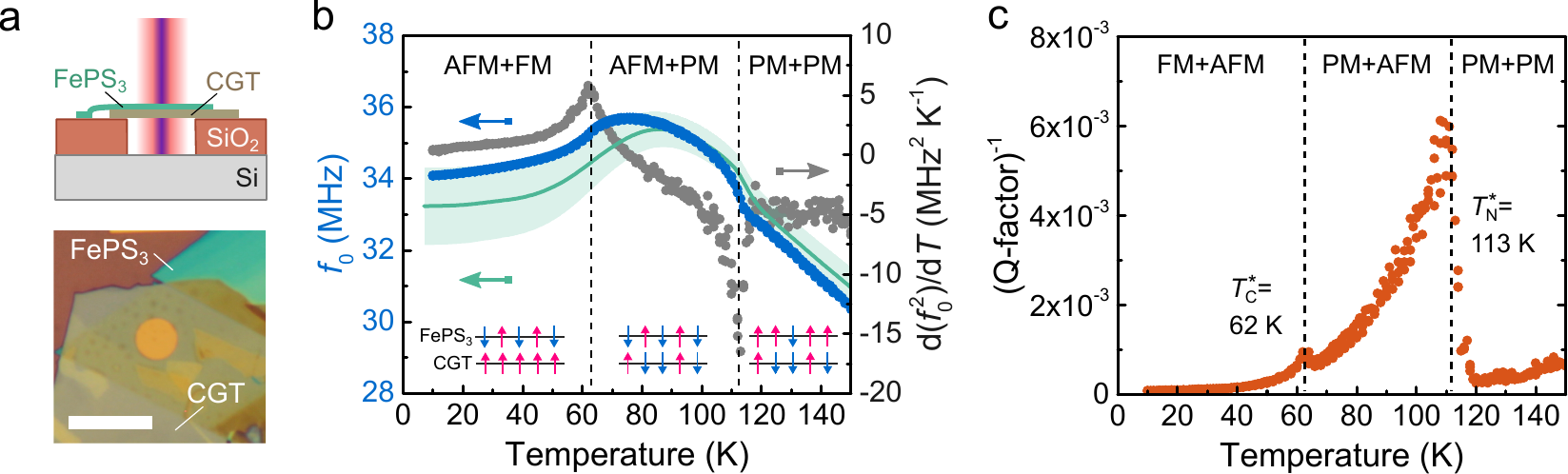}
  \caption{Mechanical properties of a CGT/FePS$_3$ ($19.8\pm0.2$ nm/$18.0\pm0.1$ nm) heterostructure membrane with a radius $r=2.5$ $\mu$m. \textbf{a} Top panel: A schematic cross-section of the suspended heterostructure membrane. Bottom panel: The optical image of the sample with the CGT/FePS$_3$ heterostructure. Scale bar: $10$ $\mu$m. \textbf{b} Filled blue circles - the measured resonance frequency as a function of temperature. Filled grey circles — the temperature derivative of $f^2_0$. Solid green line - the heterostructure resonance frequency model as described by equation~(\ref{eq_f_hetero}). The light green region indicates the higher and lower boundary of the model allowed due to the uncertainties in $h_{1,2}$, $E_{\rm CGT}$ and $f^2_0(T_0)$. The insets: Schematic pictures of a magnetic spin configuration in FePS$_3$ and CGT layers in corresponding combinations of antiferromagnetic (AFM), ferromagnetic (FM) and paramagnetic (PM) phases. \textbf{c} The mechanical damping $Q^{-1}$ as a function temperature, in which the filled orange circles represent the measured data of the CGT/FePS$_3$ heterostructure. The black vertical dashed lines in (b) and (c) represent the measured $T^*_{\rm C}=62\pm1$ K of CGT and $T^*_{\rm N}=113\pm1$ K of FePS$_3$ as determined from positions of extrema in $\frac{\mathrm{d}\left(f_{0}^2\right)}{\mathrm{d}T}$ in (b).}
  \label{fig:fourth}
\end{center}
\end{figure*}
We model the resonance frequency of a heterostructure $f_{0,\rm h}(T)$ by modifying equation~(\ref{eq:frequency}) to describe the observations in Fig.~\ref{fig:third}b. We describe the total thermally accumulated tension of the CGT/WSe$_2$ heterostructure as a sum of individual tensions in each layer \cite{HeteroFengYe2017,HeteroKim2018}: 
\begin{align}\label{hetero_strain}
n_{\rm th}(T)=&n_{\rm th,1}(T)+n_{\rm th,2}(T)\\=&\frac{E_{1}h_{1}}{(1-\nu_{1})}\epsilon_{\rm th,1}(T)+\frac{E_{2}h_{2}}{(1-\nu_{2})}\epsilon_{\rm th,2}(T).\nonumber
\end{align}
We assume that the slippage between the layers is negligible. Therefore, we obtain the final equation for the resonance frequency of the heterostructure as:
\begin{equation}
    f_{0,\rm h}(T)=\sqrt{\left(\frac{2.4048}{2\pi r}\right)^2\frac{1}{\rho_{1} h_{1}+\rho_{2} h_{2}}n_{\mathrm{th}}(T)+f_0^2(T_0)},
    \label{eq_f_hetero}
\end{equation}
where $n_{th}$ is given by equation~(\ref{hetero_strain}). We plot the model of equation~(\ref{eq_f_hetero}) for the CGT/WSe$_2$ heterostructure in Fig.~\ref{fig:third}b (solid green line). In doing so, we use the bulk $\alpha_{\rm CGT}(T)$ values \cite{Carteaux1995}, $\rho_{\rm CGT}=6091$ kg~m$^{-3}$ \cite{Carteaux1995}, $\nu_{\rm CGT}=0.22$ \cite{deJong2015}, and $E_{\rm CGT}=56.2\pm8.2$ GPa for CGT; $\epsilon_{\rm th, WSe_2}(T)$ extracted from the measured $f_{0,\rm WSe_2}(T)$ of the reference drum using equation~(\ref{eq:frequency}), $\rho_{\rm WSe_2}=9320$ kg~m$^{-3}$ \cite{WSe2Zhang2016, WSe2densityAgarwal1979}, $\nu_{\rm WSe_2}=0.19$ \cite{WSe2Zhang2016, WSe2nuZeng2015} and $E_{\rm WSe_2}=167.1\pm0.7$ GPa (measured for this membrane using the nanoindentation method described above, which is consistent which previous studies \cite{WSe2Zhang2016}) for WSe$_2$; and $f_0(94 \,\rm K)=27.2\pm0.4$ MHz. The resulting model reproduces the experiment qualitatively, yet lacks quantitative agreement above $T_{\rm C}$, most likely due to the overestimation of $\alpha_{\rm CGT}(T)$ for thin layers of CGT in contrast to its bulk value \cite{Carteaux1995}.

\subsection*{Magnetic heterostructures.} The presented methodology is not limited exclusively to the use of WSe$_2$, given that the thermal expansion coefficient of the added material is large and positive. To explore the possibility of detecting two magnetic phase transitions in the same membrane and also the possibility of having emergent properties arising from a coupling between the two flakes, we fabricate a heterostructure membrane made of a ferromagnetic CGT covered by an antiferromagnetic FePS$_3$ layer, which exhibits positive $\alpha_{\rm FePS_3}(T)$ \cite{Siskins2020}, as shown in Fig.~\ref{fig:fourth}a. Using the methods described above, we measure the resonance frequency of this suspended structure as a function of temperature. In Fig.~\ref{fig:fourth}b, we plot experimental $f_{0}(T)$ (filled blue circles) together with its $\frac{\mathrm{d}\left(f_{0}^2(T)\right)}{\mathrm{d}T}$ (filled grey circles). The temperature derivative of $f^2_0$ shows two clear extrema, indicated by black vertical dashed lines: the first one at $T^*_{\rm C}=62\pm1$ K we attribute to the $T_{\rm C}$ of CGT \cite{DiscoveryGong2017, Carteaux1995}; the second one at $T^*_{\rm N}=113\pm1$ K corresponds to the N\'{e}el temperature, $T_{\rm N}$, of FePS$_3$ \cite{FePS3Lee2016, Siskins2020}. We plot the model of equation~(\ref{eq_f_hetero}) in Fig.~\ref{fig:fourth}b (solid green line) for the CGT/FePS$_3$ heterostructure using material parameters of CGT from literature \cite{Carteaux1995,Siskins2020,deJong2015} and $E_{\rm CGT}$ as determined from Fig.~\ref{fig:first}; $E_{\rm FePS_3}=103$ GPa, $\rho_{\rm FePS_3}=3375$ kg~m$^{-3}$, $\nu_{\rm FePS_3}=0.304$ \cite{Siskins2020, FePS3YoungsHashemi2017} for FePS$_3$; and $f_0(150 \,\rm K)=30.9\pm0.7$ MHz. The model describes the experiment well with some deviations at low temperatures similar to the ones described above in Fig.~\ref{fig:third}c. In addition, we observe two peaks in $Q^{-1}(T)$, displayed in Fig.~\ref{fig:fourth}c, that we attribute to increased thermoelastic damping near the phase transition \cite{Siskins2020, Lifshitz2000, Sun2006}. The temperatures of these peaks coincide with the extrema from Fig.~\ref{fig:fourth}b, confirming the phase transitions in CGT and FePS$_3$.
\begin{figure*}
\begin{center}
  \includegraphics[width=0.85\linewidth]{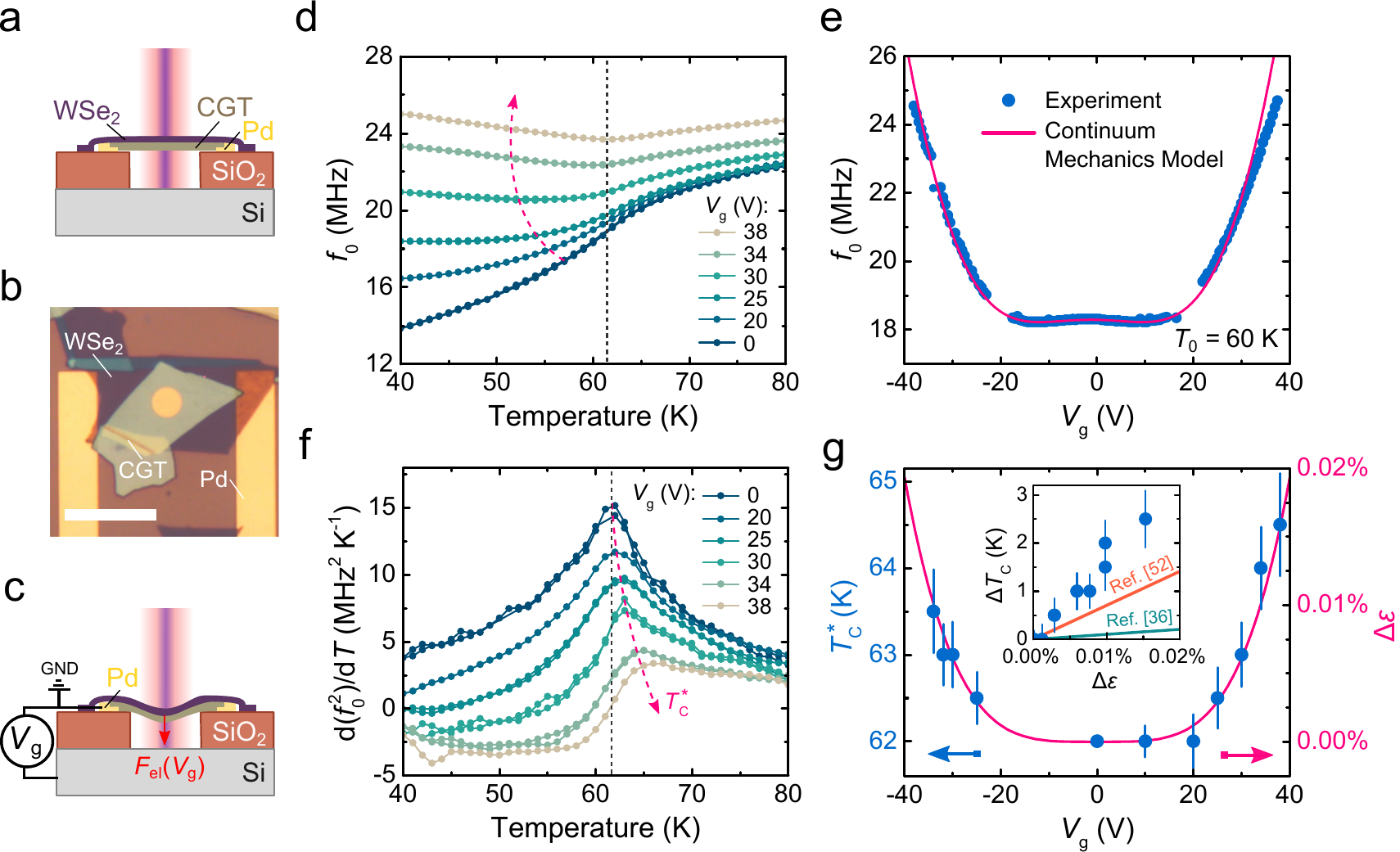}
  \caption{Curie temperature enhancement as a function of strain in a suspended CGT/WSe$_2$ heterostructure ($19.7\pm0.1$ nm/$8.5\pm0.2$ nm). \textbf{a} The schematic cross-section of the suspended CGT/WSe$_2$ heterostructure membrane. \textbf{b} The optical image of the heterostructure with materials indicated. Membrane radius: $r=4$ $\mu$m. Scale bar: $12$ $\mu$m. \textbf{c} The schematics of the electrostatic strain-tuning principle. \textbf{d} Measured resonance frequencies $f_0$ of the heterostructure membrane as a function of temperature for different gate voltages $V_{\rm g}$. \textbf{e} Filled blue circles - the measured resonance frequency as a function of $V_{\rm g}$ at $60$ K. Solid magenta line - the continuum model mechanics model fit \cite{ChenThesis2013, Weber2014}. \textbf{f} The temperature derivative of $f^2_0$ as a function of temperature for different $V_{\rm g}$. The black vertical dashed lines in (d) and (e) show the peak position in $\frac{\mathrm{d}\left(f_{0}^2\right)}{\mathrm{d}T}$ at $V_{\rm g}=0$ V. \textbf{g} Filled blue circles — the shift in measured $T^*_{\rm C}$ as a function of $V_{\rm g}$ extracted from the peak position in (e). Solid magenta line - the estimate of electrostatically induced strain $\Delta \epsilon$ as a function of $V_{\rm g}$. The inset shows $\Delta T_{\rm C}=T^*_{\rm C}(V_{\rm g})-T^*_{\rm C}(0\,\rm{V})$ as a function of added voltage induced strain $\Delta \epsilon$ in addition to calculations from Li and Yang \cite{Li2014} (solid green line) and from Dong et al \cite{StrainAnisotropyDong2020} (solid orange line). The vertical error bars in $T^*_{\rm C}$ were estimated from determining the peak position in (e) within $2 \% $ accuracy in the measured maximum.}
  \label{fig:fifth}
\end{center}
\end{figure*}
 
\subsection*{Electrostatic strain tuning of $T_{\rm C}$.} Since strong magnetostriction is responsible for the large anomalies in the mechanical response of CGT membranes at $T_{\rm C}$ \cite{Carteaux1995,MagnetostrictionTian2016} (Fig.~\ref{fig:first}), we expect that the inverse effect should also be present allowing us to tune the $T_{\rm C}$ via strain \cite{kinFaiMakJiang2020, Siskins2020}. To study this effect, we fabricate another CGT/WSe$_2$ heterostructure membrane contacted to Pd electrodes, as shown in Fig.~\ref{fig:fifth}a-c. We apply a gate voltage $V_{\rm g}$ between the heterostructure membrane and the bottom Si substrate electrode to induce an electrostatic force, $F_{\rm el}$, that pulls the membrane down and thus strains the CGT/WSe$_2$ stack. In Fig.~\ref{fig:fifth}d, we show the measured $f_0$ of the heterostructure membrane as a function of temperature for six different values of $V_{\rm g}$. A notable increase of $f_0$ as a function of $V_{\rm g}$ is evident which is attributed to the electrostatic strain introduced. To describe the gate dependence of the resonance frequency, we show the $f_0(V_{\rm g})$ relationship near the phase transition temperature at $T_0=60$ K in Fig.~\ref{fig:fifth}e. We fit the measurement data to a continuum mechanics model of a circular membrane resonator \cite{ChenThesis2013, Weber2014} (see Experimental Section), which describes the experiment well. We thus estimate the added electrostatic strain \cite{kinFaiMakJiang2020, Weber2014} as $\Delta \epsilon\approx\frac{2}{3}\left(\frac{\varepsilon_0 r}{8g_0^2n(T_0)}\right)^2V_{\rm g}^4$, where $\varepsilon_0$ is the dielectric permittivity of vacuum, and $n(T_0)=2.538\pm0.003$ Nm$^{-1}$ the total tension in the suspended heterostructure at $T_0=60$ K (extracted from the fit in Fig.~\ref{fig:fifth}e).

Apart from the effect of electrostatically induced strain on $f_0(V_{\rm g})$ in Fig.~\ref{fig:fifth}d, the characteristic feature in the $f_0(T)$ curve, that we attribute to the Curie temperature of CGT \cite{Siskins2020}, shifts to higher temperatures. This trend becomes even more apparent in Fig.~\ref{fig:fifth}f, where we plot $\frac{\mathrm{d}\left(f_{0}^2\right)}{\mathrm{d}T}$ for six different $V_{\rm g}$. We extract $T^*_{\rm C}$ from the peak positions in $\frac{\mathrm{d}\left(f_{0}^2\right)}{\mathrm{d}T}$ for multiple $V_{\rm g}$ and plot it against the gate-induced strain, $\Delta \epsilon$, in Fig.~\ref{fig:fifth}g. The observed enhancement in $T^*_{\rm C}$ qualitatively agrees with the added tensile strain dependence symmetrically both for positive and negative $V_{\rm g}$ as shown in Fig.~\ref{fig:fifth}g. This indicates that the gate-induced strain has a dominant contribution to the change in the $T^*_{\rm C}(V_{\rm g})$ instead of a field effect \cite{CGTGateAnisotropyVerzhbitskiy2020,EFieldWang2018}. In the inset of Fig.~\ref{fig:fifth}g, we plot the experimental $\Delta T_{\rm C}=T^*_{\rm C}(V_{\rm g})-T^*_{\rm C}(0\,\rm V)$ versus the estimated strain $\Delta \epsilon$, together with predictions from density-functional theory calculations for 2D CGT by Li and Yang \cite{Li2014}, considering the direct exchange interaction (solid green line), and by Dong et al \cite{StrainAnisotropyDong2020}, attributing the enhancement of $T_{\rm C}$ to the super-exchange interaction via Cr-Te-Cr bonds (solid orange line). The closer experimental agreement with the results of Dong et al. \cite{StrainAnisotropyDong2020} suggests that it is the decreasing energy difference between $3d$ orbitals of Cr and $5p$ orbitals of Te in nearly $\sim90^{\circ}$ Cr-Te-Cr bonds of CGT that is likely to contribute to the enhancement of $T^*_{\rm C}$ as a function of strain \cite{StrainAnisotropyDong2020}. Interestingly, we observed that the maximal $T^*_{\rm C}$ increase by mechanical strain was as much as $2.5\pm0.6$ K at $\Delta \epsilon\approx0.016\%$. This is comparable to what was previously achieved in bulk CGT by applying an out-of-plane magnetic field of $0.7-1$ T \cite{DiscoveryGong2017,HFieldSelter2020}, proving that strain-based nanomechanical methods provide a compelling route for controlling and probing magneto-elastic coupling in ferromagnetic 2D layers and heterostructures. 

\section*{Discussion}
In conclusion, we have probed the mechanical properties of CGT membranes by both static and dynamic nanomechanical means. We demonstrated that due to CGT's strong magnetostriction \cite{Carteaux1995, MagnetostrictionTian2016} and large negative thermal expansion near $T_{\rm C}$, bare CGT membranes experience strong resonance frequency reductions near their ferromagnetic Curie temperature and an accumulation of compressive strain. The compressive strain can produce buckling, wrinkling or sagging of the CGT layers, that significantly complicates the analysis for nanomechanical measurements of $T_{\rm C}$ \cite{Siskins2020}. We showed that this issue can be solved by integrating CGT in heterostructures with other materials with positive thermal expansion coefficients. This methodology can also be applied using materials that have phase transitions themselves, which we demonstrate by probing both $T_{\rm N}$ of FePS$_3$ and $T_{\rm C}$ of CGT within a single measurement of a CGT/FePS$_3$ heterostructure. The presented measurements and methodologies for nanomechanical characterisation of magnetic heterostructures open up possibilities to investigate magnetic properties that are the result of interfacial interactions in 2D magnetic heterostructures membranes \cite{ExchangeBiasZhang2020,InterfaceWang2020}, allowing also novel magnetic properties to be discovered near the monolayer limit while excluding substrate interactions. In addition, we expect that focused studies of the mechanical dissipation in such heterostructures as a function of temperature and strain could in the future provide more insights to thermodynamics and heat transport in 2D materials involved \cite{Siskins2020, TauDolleman2017, TauMorell2019}. Finally, we showed that $T_{\rm C}$ of CGT can be enhanced by gate-induced electrostatic straining, demonstrating control of magnetic order in these ferromagnetic heterostructures using strain. We anticipate that these studies in the future will lead to the development of membrane devices with gate-controlled magnetic actuation \cite{MagneticExitationBenShimon2021} for low-power spintronic applications. 

\section*{Methods}

\paragraph*{\textbf{Device fabrication and characterization.}}
We first pattern a diced Si/SiO$_2$ wafer and define circular holes with a radius of $r=1-2.5$ $\mu$m and cavity depth of $g_0=285$ nm using an e-beam lithography and reactive ion etching. SiO$_2$ provides an electrical insulation between subsequently transferred 2D material membranes and the bottom Si electrode. To realize electrical contact to some samples for electrostatic experiments, Pd electrodes are pre-patterned on top of Si/SiO$_2$ chips by a lift-off technique. To create suspended membranes, thin flakes of CGT, FePS$_3$ and WSe$_2$ crystals are mechanically exfoliated and transferred on a pre-patterned chip by the all-dry viscoelastic stamping method \cite{CastellanosGomez2014} directly after exfoliation. Flakes of van der Waals crystals are exfoliated from high quality synthetically grown crystals with known stoichiometry. Using the same method, flakes are deterministically stacked on top of each other forming heterostructures. We stack FePS$_3$ and WSe$_2$ flakes on top of CGT to minimise the exposure time of the suspended CGT layers to air. Subsequently, samples were kept in an oxygen-free environment to avoid degradation.

\paragraph*{\textbf{Atomic force microscopy characterisation.}}
Atomic force microscopy height profile scans and inspection are performed in tapping mode on a Bruker Dimension FastScan AFM. We use cantilevers with spring constants of $k_{\rm c}=30-40$ Nm$^{-1}$ for inspection. Error bars on reported thickness values are determined by measuring multiple profile scans of the same flake. For force nanoindentation experiments, we use two different cantilevers with spring constants of $k_{\rm c} = 8.87\pm0.08$ Nm$^{-1}$ and $k_{\rm c} = 18.90\pm0.11$ Nm$^{-1}$. $k_{\rm c}$ for each cantilever was calibrated using thermal and solid surface deflection. We use cantilever tips with a tip radius of $\sim7-10$ nm, as confirmed by scanning electron microscope imaging.

\paragraph*{\textbf{Laser interferometry measurements.}}
The sample is mounted on a $xy$ piezopositioning stage inside a dry optical $4$ K cryostat Montana Instruments Cryostation s50. A local sample heater is used to perform temperature sweeps at a rate of $\sim3$ K per minute, keeping the pressure in the chamber below $10^{-6}$ mbar. A power modulated blue diode laser of $\lambda=405$ nm is used to optothermally excite the motion of the membrane. Membrane displacement is then measured using a laser interferometer with a He$-$Ne laser beam of $\lambda=632$ nm focused on the suspended membrane. In doing so, the interfering reflections from the membrane and the Si electrode underneath are recorded using a photodiode and processed by a vector network analyzer. All measurements were performed at incident laser powers of $P_{\rm{red}}\leq8$ $\mu$W and $P_{\rm{blue}}\leq1$ $\mu$W. Laser spot size is on the order of $\sim1$ $\mu$m. During the data acquisition it is checked for all membranes that resonance frequency changes due to laser heating are insignificant. During data acquisition temperature is kept constant with $\sim10$ mK stability by the local heater and a closed feedback loop controller.

\paragraph*{\textbf{Continuum mechanics model of electrostatically strained circular membrane}}
A membrane suspended over a circular cavity forms a capacitor with a bottom gate electrode underneath. The change in gate voltage causes the membrane to deflect, tuning the tension and producing a shift in the resonance frequency. The resonance frequency of the fundamental vibration mode of the drum as a function of gate voltage is described as \cite{ChenThesis2013}: 
\begin{equation}
\begin{split}
    f_0&(V_{\rm g})=\\
    &\frac{1}{2\pi}\sqrt{\frac{1}{m_{\rm{eff}}}\left[\frac{2\pi Eh\epsilon_0}{1-\nu^2}+\frac{8\pi Eh}{\left(1-\nu^2\right)r^2}z_{\rm g}^2-\frac{1}{2}\frac{\partial^2 C_{\rm g}}{\partial z^2}V_{\rm g}^2\right]},
\end{split}
    \label{eq:springconstant}
\end{equation}
where $z_{\rm g}\approx\frac{\varepsilon_0r^2}{8g_{0}^2n(T_0)}V_{\rm g}^2$ is the maximal deflection at the membrane centre \cite{Weber2014}, $\frac{\partial^2C_{\rm g}}{\partial z^2}\approx \frac{0.542 \varepsilon_0 \pi r^2}{g_{0}^3}$ the second derivative of capacitance \cite{Weber2014,GorbachevWill2017}, and $m_{\rm{eff}}=0.27\rho h \pi r^2$ the effective mass. In the case of a heterostructure membrane, material parameters are substituted for the heterostructure analogues \cite{HeteroFengYe2017}: $E=\frac{E_{1}h_{1}+E_{2}h_{2}}{h_{1}+h_{2}}$, $\rho=\frac{\rho_{1}h_{1}+\rho_{2}h_{2}}{h_{1}+h_{2}}$, $\nu=\frac{\nu_{1}h_{1}+\nu_{2}h_{2}}{h_{1}+h_{2}}$, and $h=h_{1}+h_{2}$. For the CGT/WSe$_2$ heterostructure ($19.7\pm0.1$ nm/$8.5\pm0.2$ nm) in Fig.~\ref{fig:fifth} we use the following material parameters: $E=89.6$ GPa, $h=28.2$ nm, $\nu=0.21 $, $\rho=7064$ kg~m$^{-3}$. We extract  $\epsilon=0.079\pm0.004\%$, $\frac{\partial^2 C_{\rm g}}{\partial z^2}=3.46\pm0.11$ mFm$^{-2}$ and $m_{\rm eff}=996.45$ fg from the fit in Fig.~\ref{fig:fifth}e, that are close to the expected values \cite{Weber2014,GorbachevWill2017} of $\frac{\partial^2 C_{\rm g}}{\partial z^2}=2.61$ mFm$^{-2}$ and $m_{\rm eff}=675.9$ fg.

 
\textbf{Acknowledgements} \par 
The authors thank B. Simon for continuous support and feedback on the manuscript. M.\v{S}., M.L., H.S.J.v.d.Z. and P.G.S. acknowledge funding from the European Union’s Horizon 2020 research and innovation program under grant agreement number 881603. H.S.J.v.d.Z., E.C. and S.M.-V. thank COST Action MOLSPIN CA15128; E.C. and S.M.-V. acknowledge the financial support from the European Union (ERC AdG Mol-2D 788222), the Spanish MICINN (MAT2017-89993-R co-financed by FEDER and Excellence Unit "Mar\'{i}a de Maeztu", CEX2019-000919-M) and the Generalitat Valenciana (Prometeo program and PO FEDER Program, ref. IDIFEDER/2018/061 and IDIFEDER/2020/063). W.X., S.J., and W.H. thank National Natural Science Foundation of China (11974025).

 
\textbf{Author Contributions} \par 
M.\v{S}., S.K., E.C., H.S.J.v.d.Z., T.v.d.S. and P.G.S. conceived the experiments. M.\v{S}. performed the
laser interferometry measurements. M.\v{S}. and M.L. fabricated and inspected the samples. S.M.-V. and E.C. synthesized and characterized the FePS$_3$ crystals. W.X., S.J. and W.H. synthesized and characterized the Cr$_2$Ge$_2$Te$_6$ crystals. M.\v{S}., S.K. and B.J.M.S. analyzed and modeled the experimental data. H.S.J.v.d.Z., T.v.d.S. and P.G.S. supervised the project. The paper was jointly written by all authors with a main contribution from M.\v{S}. All authors discussed the results and commented on the paper.

\section*{References}

\begin{thebibliography}{10}
\expandafter\ifx\csname url\endcsname\relax
  \def\url#1{\texttt{#1}}\fi
\expandafter\ifx\csname urlprefix\endcsname\relax\def\urlprefix{URL }\fi
\providecommand{\bibinfo}[2]{#2}
\providecommand{\eprint}[2][]{\url{#2}}

\bibitem{CrI3Huang2017}
\bibinfo{author}{Huang, B.} \emph{et~al.}
\newblock \bibinfo{title}{Layer-dependent ferromagnetism in a van der {Waals}
  crystal down to the monolayer limit}.
\newblock \emph{\bibinfo{journal}{Nature}} \textbf{\bibinfo{volume}{546}},
  \bibinfo{pages}{270} (\bibinfo{year}{2017}).

\bibitem{DiscoveryGong2017}
\bibinfo{author}{Gong, C.} \emph{et~al.}
\newblock \bibinfo{title}{Discovery of intrinsic ferromagnetism in
  two-dimensional van der {Waals} crystals}.
\newblock \emph{\bibinfo{journal}{Nature}} \textbf{\bibinfo{volume}{546}},
  \bibinfo{pages}{265} (\bibinfo{year}{2017}).

\bibitem{FePS3Lee2016}
\bibinfo{author}{Lee, J.-U.} \emph{et~al.}
\newblock \bibinfo{title}{{Ising}-type magnetic ordering in atomically thin
  {FePS$_3$}}.
\newblock \emph{\bibinfo{journal}{Nano Lett.}} \textbf{\bibinfo{volume}{16}},
  \bibinfo{pages}{7433} (\bibinfo{year}{2016}).

\bibitem{Gibertini2019}
\bibinfo{author}{Gibertini, M.}, \bibinfo{author}{Koperski, M.},
  \bibinfo{author}{Morpurgo, A.~F.} \& \bibinfo{author}{Novoselov, K.~S.}
\newblock \bibinfo{title}{Magnetic {2D} materials and heterostructures}.
\newblock \emph{\bibinfo{journal}{Nat. Nanotechnol.}}
  \textbf{\bibinfo{volume}{14}}, \bibinfo{pages}{408} (\bibinfo{year}{2019}).

\bibitem{Carteaux1995}
\bibinfo{author}{Carteaux, V.}, \bibinfo{author}{Brunet, D.},
  \bibinfo{author}{Ouvrard, G.} \& \bibinfo{author}{Andre, G.}
\newblock \bibinfo{title}{Crystallographic, magnetic and electronic structures
  of a new layered ferromagnetic compound {Cr$_2$Ge$_2$Te$_6$}}.
\newblock \emph{\bibinfo{journal}{J. Phys. Condens. Matter.}}
  \textbf{\bibinfo{volume}{7}}, \bibinfo{pages}{69} (\bibinfo{year}{1995}).

\bibitem{CGTGateAnisotropyVerzhbitskiy2020}
\bibinfo{author}{Verzhbitskiy, I.~A.} \emph{et~al.}
\newblock \bibinfo{title}{Controlling the magnetic anisotropy in
  {Cr$_2$Ge$_2$Te$_6$} by electrostatic gating}.
\newblock \emph{\bibinfo{journal}{Nat. Electron.}}
  \textbf{\bibinfo{volume}{3}}, \bibinfo{pages}{460} (\bibinfo{year}{2020}).

\bibitem{EFieldWang2018}
\bibinfo{author}{Wang, Z.} \emph{et~al.}
\newblock \bibinfo{title}{Electric-field control of magnetism in a few-layered
  van der {Waals} ferromagnetic semiconductor}.
\newblock \emph{\bibinfo{journal}{Nat. Nanotechnol.}}
  \textbf{\bibinfo{volume}{13}}, \bibinfo{pages}{554} (\bibinfo{year}{2018}).

\bibitem{HFieldSelter2020}
\bibinfo{author}{Selter, S.}, \bibinfo{author}{Bastien, G.},
  \bibinfo{author}{Wolter, A. U.~B.}, \bibinfo{author}{Aswartham, S.} \&
  \bibinfo{author}{B\"{u}chner, B.}
\newblock \bibinfo{title}{Magnetic anisotropy and low-field magnetic phase
  diagram of the quasi-two-dimensional ferromagnet {Cr$_2$Ge$_2$Te$_6$}}.
\newblock \emph{\bibinfo{journal}{Phys. Rev. B}}
  \textbf{\bibinfo{volume}{101}}, \bibinfo{pages}{014440}
  (\bibinfo{year}{2020}).

\bibitem{PressureSun2018}
\bibinfo{author}{Sun, Y.} \emph{et~al.}
\newblock \bibinfo{title}{Effects of hydrostatic pressure on spin-lattice
  coupling in two-dimensional ferromagnetic {Cr$_2$Ge$_2$Te$_6$}}.
\newblock \emph{\bibinfo{journal}{Appl. Phys. Lett.}}
  \textbf{\bibinfo{volume}{112}}, \bibinfo{pages}{072409}
  (\bibinfo{year}{2018}).

\bibitem{PressureSakurai2021}
\bibinfo{author}{Sakurai, T.} \emph{et~al.}
\newblock \bibinfo{title}{Pressure control of the magnetic anisotropy of the
  quasi-two-dimensional van der {Waals} ferromagnet {Cr$_2$Ge$_2$Te$_6$}}.
\newblock \emph{\bibinfo{journal}{Phys. Rev. B}}
  \textbf{\bibinfo{volume}{103}}, \bibinfo{pages}{024404}
  (\bibinfo{year}{2021}).

\bibitem{IonsWang2019}
\bibinfo{author}{Wang, N.} \emph{et~al.}
\newblock \bibinfo{title}{Transition from ferromagnetic semiconductor to
  ferromagnetic metal with enhanced curie temperature in {Cr$_2$Ge$_2$Te$_6$}
  via organic ion intercalation}.
\newblock \emph{\bibinfo{journal}{J. Am. Chem. Soc.}}
  \textbf{\bibinfo{volume}{141}}, \bibinfo{pages}{17166}
  (\bibinfo{year}{2019}).

\bibitem{SpinOrbitGupta2020}
\bibinfo{author}{Gupta, V.} \emph{et~al.}
\newblock \bibinfo{title}{Manipulation of the van der {Waals} magnet
  {Cr$_2$Ge$_2$Te$_6$} by spin{\textendash}orbit torques}.
\newblock \emph{\bibinfo{journal}{Nano Lett.}} \textbf{\bibinfo{volume}{20}},
  \bibinfo{pages}{7482} (\bibinfo{year}{2020}).

\bibitem{SpinOrbitOstwal2020}
\bibinfo{author}{Ostwal, V.}, \bibinfo{author}{Shen, T.} \&
  \bibinfo{author}{Appenzeller, J.}
\newblock \bibinfo{title}{Efficient spin-orbit torque switching of the
  semiconducting van der {Waals} ferromagnet {Cr$_2$Ge$_2$Te$_6$}}.
\newblock \emph{\bibinfo{journal}{Adv. Mater.}} \textbf{\bibinfo{volume}{32}},
  \bibinfo{pages}{1906021} (\bibinfo{year}{2020}).

\bibitem{MagnetostrictionTian2016}
\bibinfo{author}{Tian, Y.}, \bibinfo{author}{Gray, M.~J.}, \bibinfo{author}{Ji,
  H.}, \bibinfo{author}{Cava, R.~J.} \& \bibinfo{author}{Burch, K.~S.}
\newblock \bibinfo{title}{Magneto-elastic coupling in a potential ferromagnetic
  {2D} atomic crystal}.
\newblock \emph{\bibinfo{journal}{2D Mater.}} \textbf{\bibinfo{volume}{3}},
  \bibinfo{pages}{025035} (\bibinfo{year}{2016}).

\bibitem{Kozlenko2021}
\bibinfo{author}{Kozlenko, D.~P.} \emph{et~al.}
\newblock \bibinfo{title}{Spin-induced negative thermal expansion and
  spin{\textendash}phonon coupling in van der {Waals} material {CrBr$_3$}}.
\newblock \emph{\bibinfo{journal}{npj Quantum Mater.}}
  \textbf{\bibinfo{volume}{6}}, \bibinfo{pages}{19} (\bibinfo{year}{2021}).

\bibitem{Casto2015}
\bibinfo{author}{Casto, L.~D.} \emph{et~al.}
\newblock \bibinfo{title}{Strong spin-lattice coupling in {CrSiTe$_3$}}.
\newblock \emph{\bibinfo{journal}{{APL} Mater.}} \textbf{\bibinfo{volume}{3}},
  \bibinfo{pages}{041515} (\bibinfo{year}{2015}).

\bibitem{McGuire2015}
\bibinfo{author}{McGuire, M.~A.}, \bibinfo{author}{Dixit, H.},
  \bibinfo{author}{Cooper, V.~R.} \& \bibinfo{author}{Sales, B.~C.}
\newblock \bibinfo{title}{Coupling of crystal structure and magnetism in the
  layered, ferromagnetic insulator {CrI$_3$}}.
\newblock \emph{\bibinfo{journal}{Chem. Mater.}} \textbf{\bibinfo{volume}{27}},
  \bibinfo{pages}{612--620} (\bibinfo{year}{2015}).

\bibitem{CrCl3NTESchneeloch2021}
\bibinfo{author}{Schneeloch, J.~A.} \emph{et~al.}
\newblock \bibinfo{title}{Gapless {D}irac magnons in {CrCl$_3$}}
  (\bibinfo{year}{2021}).
\newblock \bibinfo{note}{Preprint at https://arxiv.org/abs/2110.10771v1}.

\bibitem{StraintronicsMiao2021}
\bibinfo{author}{Miao, F.}, \bibinfo{author}{Liang, S.-J.} \&
  \bibinfo{author}{Cheng, B.}
\newblock \bibinfo{title}{Straintronics with van der {Waals} materials}.
\newblock \emph{\bibinfo{journal}{npj Quantum Mater.}}
  \textbf{\bibinfo{volume}{6}}, \bibinfo{pages}{59} (\bibinfo{year}{2021}).

\bibitem{StrainYang2021}
\bibinfo{author}{Yang, S.}, \bibinfo{author}{Chen, Y.} \&
  \bibinfo{author}{Jiang, C.}
\newblock \bibinfo{title}{Strain engineering of two-dimensional materials:
  Methods, properties, and applications}.
\newblock \emph{\bibinfo{journal}{{InfoMat}}} \textbf{\bibinfo{volume}{3}},
  \bibinfo{pages}{397--420} (\bibinfo{year}{2021}).

\bibitem{GrapheneResonatorChen2009}
\bibinfo{author}{Chen, C.} \emph{et~al.}
\newblock \bibinfo{title}{Performance of monolayer graphene nanomechanical
  resonators with electrical readout}.
\newblock \emph{\bibinfo{journal}{Nat. Nanotechnol.}}
  \textbf{\bibinfo{volume}{4}}, \bibinfo{pages}{861} (\bibinfo{year}{2009}).

\bibitem{DynamicStrainZhang2020}
\bibinfo{author}{Zhang, X.} \emph{et~al.}
\newblock \bibinfo{title}{Dynamically-enhanced strain in atomically thin
  resonators}.
\newblock \emph{\bibinfo{journal}{Nat. Commun.}} \textbf{\bibinfo{volume}{11}},
  \bibinfo{pages}{5526} (\bibinfo{year}{2020}).

\bibitem{GrapheneTunableOscillatorChen2013}
\bibinfo{author}{Chen, C.} \emph{et~al.}
\newblock \bibinfo{title}{Graphene mechanical oscillators with tunable
  frequency}.
\newblock \emph{\bibinfo{journal}{Nat. Nanotechnol.}}
  \textbf{\bibinfo{volume}{8}}, \bibinfo{pages}{923} (\bibinfo{year}{2013}).

\bibitem{SensorsLemme2020}
\bibinfo{author}{Lemme, M.~C.} \emph{et~al.}
\newblock \bibinfo{title}{Nanoelectromechanical sensors based on suspended {2D}
  materials}.
\newblock \emph{\bibinfo{journal}{Research}} \textbf{\bibinfo{volume}{2020}},
  \bibinfo{pages}{8748602} (\bibinfo{year}{2020}).

\bibitem{Lee2008}
\bibinfo{author}{Lee, C.}, \bibinfo{author}{Wei, X.}, \bibinfo{author}{Kysar,
  J.~W.} \& \bibinfo{author}{Hone, J.}
\newblock \bibinfo{title}{Measurement of the elastic properties and intrinsic
  strength of monolayer graphene}.
\newblock \emph{\bibinfo{journal}{Science}} \textbf{\bibinfo{volume}{321}},
  \bibinfo{pages}{5887} (\bibinfo{year}{2008}).

\bibitem{Siskins2020}
\bibinfo{author}{{\v{S}}i{\v{s}}kins, M.} \emph{et~al.}
\newblock \bibinfo{title}{Magnetic and electronic phase transitions probed by
  nanomechanical resonators}.
\newblock \emph{\bibinfo{journal}{Nat. Commun.}} \textbf{\bibinfo{volume}{11}},
  \bibinfo{pages}{2698} (\bibinfo{year}{2020}).

\bibitem{Morell2016}
\bibinfo{author}{Morell, N.} \emph{et~al.}
\newblock \bibinfo{title}{High quality factor mechanical resonators based on
  {WSe$_2$} monolayers}.
\newblock \emph{\bibinfo{journal}{Nano Lett.}} \textbf{\bibinfo{volume}{16}},
  \bibinfo{pages}{5102} (\bibinfo{year}{2016}).

\bibitem{HeteroFengYe2017}
\bibinfo{author}{Ye, F.}, \bibinfo{author}{Lee, J.} \& \bibinfo{author}{Feng,
  P. X.-L.}
\newblock \bibinfo{title}{Atomic layer {MoS$_2$}-graphene van der {Waals}
  heterostructure nanomechanical resonators}.
\newblock \emph{\bibinfo{journal}{Nanoscale}} \textbf{\bibinfo{volume}{9}},
  \bibinfo{pages}{18208} (\bibinfo{year}{2017}).

\bibitem{kinFaiMakJiang2020}
\bibinfo{author}{Jiang, S.}, \bibinfo{author}{Xie, H.}, \bibinfo{author}{Shan,
  J.} \& \bibinfo{author}{Mak, K.~F.}
\newblock \bibinfo{title}{Exchange magnetostriction in two-dimensional
  antiferromagnets}.
\newblock \emph{\bibinfo{journal}{Nat. Mater.}} \textbf{\bibinfo{volume}{19}},
  \bibinfo{pages}{1295} (\bibinfo{year}{2020}).

\bibitem{NeelVectorNi2021}
\bibinfo{author}{Ni, Z.} \emph{et~al.}
\newblock \bibinfo{title}{Imaging the {N}{\'{e}}el vector switching in the
  monolayer antiferromagnet {MnPSe$_3$} with strain-controlled {Ising} order}.
\newblock \emph{\bibinfo{journal}{Nat. Nanotechnol.}}
  \textbf{\bibinfo{volume}{16}}, \bibinfo{pages}{782--787}
  (\bibinfo{year}{2021}).

\bibitem{HeteroLiu2014}
\bibinfo{author}{Liu, K.} \emph{et~al.}
\newblock \bibinfo{title}{Elastic properties of chemical-vapor-deposited
  monolayer {MoS$_2$}, {WS$_2$}, and their bilayer heterostructures}.
\newblock \emph{\bibinfo{journal}{Nano Lett.}} \textbf{\bibinfo{volume}{14}},
  \bibinfo{pages}{5097} (\bibinfo{year}{2014}).

\bibitem{HeteroKim2018}
\bibinfo{author}{Kim, S.}, \bibinfo{author}{Yu, J.} \& \bibinfo{author}{van~der
  Zande, A.~M.}
\newblock \bibinfo{title}{Nano-electromechanical drumhead resonators from
  two-dimensional material bimorphs}.
\newblock \emph{\bibinfo{journal}{Nano Lett.}} \textbf{\bibinfo{volume}{18}},
  \bibinfo{pages}{6686} (\bibinfo{year}{2018}).

\bibitem{HeterostructureNovoselov2016}
\bibinfo{author}{Novoselov, K.~S.}, \bibinfo{author}{Mishchenko, A.},
  \bibinfo{author}{Carvalho, A.} \& \bibinfo{author}{Neto, A. H.~C.}
\newblock \bibinfo{title}{{2D} materials and van der {Waals} heterostructures}.
\newblock \emph{\bibinfo{journal}{Science}} \textbf{\bibinfo{volume}{353}},
  \bibinfo{pages}{aac9439} (\bibinfo{year}{2016}).

\bibitem{CastellanosGomez2012}
\bibinfo{author}{Castellanos-Gomez, A.} \emph{et~al.}
\newblock \bibinfo{title}{Elastic properties of freely suspended
  {MoS$_2$}nanosheets}.
\newblock \emph{\bibinfo{journal}{Adv. Mater.}} \textbf{\bibinfo{volume}{24}},
  \bibinfo{pages}{772} (\bibinfo{year}{2012}).

\bibitem{deJong2015}
\bibinfo{author}{de~Jong, M.} \emph{et~al.}
\newblock \bibinfo{title}{Charting the complete elastic properties of inorganic
  crystalline compounds}.
\newblock \emph{\bibinfo{journal}{Sci. Data}} \textbf{\bibinfo{volume}{2}},
  \bibinfo{pages}{150009} (\bibinfo{year}{2015}).

\bibitem{Li2014}
\bibinfo{author}{Li, X.} \& \bibinfo{author}{Yang, J.}
\newblock \bibinfo{title}{{CrXTe$_3$ (X = Si, Ge)} nanosheets: two dimensional
  intrinsic ferromagnetic semiconductors}.
\newblock \emph{\bibinfo{journal}{J. Mater. Chem. C}}
  \textbf{\bibinfo{volume}{2}}, \bibinfo{pages}{7071} (\bibinfo{year}{2014}).

\bibitem{CrI3CantosPrieto2021}
\bibinfo{author}{Cantos-Prieto, F.} \emph{et~al.}
\newblock \bibinfo{title}{Layer-dependent mechanical properties and enhanced
  plasticity in the van der {Waals} chromium trihalide magnets}.
\newblock \emph{\bibinfo{journal}{Nano Lett.}} \textbf{\bibinfo{volume}{21}},
  \bibinfo{pages}{3379--3385} (\bibinfo{year}{2021}).

\bibitem{MoS2ResonatorCastellanosGomez2013}
\bibinfo{author}{Castellanos-Gomez, A.} \emph{et~al.}
\newblock \bibinfo{title}{Single-layer {MoS$_2$} mechanical resonators}.
\newblock \emph{\bibinfo{journal}{Adv. Mater.}} \textbf{\bibinfo{volume}{25}},
  \bibinfo{pages}{6719} (\bibinfo{year}{2013}).

\bibitem{Lyon1977}
\bibinfo{author}{Lyon, K.~G.}, \bibinfo{author}{Salinger, G.~L.},
  \bibinfo{author}{Swenson, C.~A.} \& \bibinfo{author}{White, G.~K.}
\newblock \bibinfo{title}{Linear thermal expansion measurements on silicon from
  6 to 340 {K}}.
\newblock \emph{\bibinfo{journal}{J. Appl. Phys.}}
  \textbf{\bibinfo{volume}{48}}, \bibinfo{pages}{865} (\bibinfo{year}{1977}).

\bibitem{MechanicalMemoryChen2021}
\bibinfo{author}{Chen, T.}, \bibinfo{author}{Pauly, M.} \&
  \bibinfo{author}{Reis, P.~M.}
\newblock \bibinfo{title}{A reprogrammable mechanical metamaterial with stable
  memory}.
\newblock \emph{\bibinfo{journal}{Nature}} \textbf{\bibinfo{volume}{589}},
  \bibinfo{pages}{386} (\bibinfo{year}{2021}).

\bibitem{BistableMahboob2008}
\bibinfo{author}{Mahboob, I.} \& \bibinfo{author}{Yamaguchi, H.}
\newblock \bibinfo{title}{Bit storage and bit flip operations in an
  electromechanical oscillator}.
\newblock \emph{\bibinfo{journal}{Nat. Nanotechnol.}}
  \textbf{\bibinfo{volume}{3}}, \bibinfo{pages}{275} (\bibinfo{year}{2008}).

\bibitem{BistableRoodenburg2009}
\bibinfo{author}{Roodenburg, D.}, \bibinfo{author}{Spronck, J.~W.},
  \bibinfo{author}{van~der Zant, H. S.~J.} \& \bibinfo{author}{Venstra, W.~J.}
\newblock \bibinfo{title}{Buckling beam micromechanical memory with on-chip
  readout}.
\newblock \emph{\bibinfo{journal}{Appl. Phys. Lett.}}
  \textbf{\bibinfo{volume}{94}}, \bibinfo{pages}{183501}
  (\bibinfo{year}{2009}).

\bibitem{WSe2Zhang2016}
\bibinfo{author}{Zhang, R.}, \bibinfo{author}{Koutsos, V.} \&
  \bibinfo{author}{Cheung, R.}
\newblock \bibinfo{title}{Elastic properties of suspended multilayer
  {WSe$_2$}}.
\newblock \emph{\bibinfo{journal}{Appl. Phys. Lett.}}
  \textbf{\bibinfo{volume}{108}}, \bibinfo{pages}{042104}
  (\bibinfo{year}{2016}).

\bibitem{WSe2Cakir2014}
\bibinfo{author}{{\c{C}}ak{\i}r, D.}, \bibinfo{author}{Peeters, F.~M.} \&
  \bibinfo{author}{Sevik, C.}
\newblock \bibinfo{title}{Mechanical and thermal properties of h-{MX$_2$ (M =
  Cr, Mo, W; X = O, S, Se, Te)} monolayers: A comparative study}.
\newblock \emph{\bibinfo{journal}{Appl. Phys. Lett.}}
  \textbf{\bibinfo{volume}{104}}, \bibinfo{pages}{203110}
  (\bibinfo{year}{2014}).

\bibitem{Lifshitz2000}
\bibinfo{author}{Lifshitz, R.} \& \bibinfo{author}{Roukes, M.~L.}
\newblock \bibinfo{title}{Thermoelastic damping in micro- and nanomechanical
  systems}.
\newblock \emph{\bibinfo{journal}{Phys. Rev. B}} \textbf{\bibinfo{volume}{61}},
  \bibinfo{pages}{5600} (\bibinfo{year}{2000}).

\bibitem{Sun2006}
\bibinfo{author}{Sun, Y.}, \bibinfo{author}{Fang, D.} \& \bibinfo{author}{Soh,
  A.~K.}
\newblock \bibinfo{title}{Thermoelastic damping in micro-beam resonators}.
\newblock \emph{\bibinfo{journal}{Int. J. Solids. Struct.}}
  \textbf{\bibinfo{volume}{43}}, \bibinfo{pages}{3213} (\bibinfo{year}{2006}).

\bibitem{WSe2densityAgarwal1979}
\bibinfo{author}{Agarwal, M.} \& \bibinfo{author}{Wani, P.}
\newblock \bibinfo{title}{Growth conditions and crystal structure parameters of
  layer compounds in the series {Mo$_{1-x}$W$_x$Se$_2$}}.
\newblock \emph{\bibinfo{journal}{Mater. Res. Bull.}}
  \textbf{\bibinfo{volume}{14}}, \bibinfo{pages}{825} (\bibinfo{year}{1979}).

\bibitem{WSe2nuZeng2015}
\bibinfo{author}{Zeng, F.}, \bibinfo{author}{Zhang, W.-B.} \&
  \bibinfo{author}{Tang, B.-Y.}
\newblock \bibinfo{title}{Electronic structures and elastic properties of
  monolayer and bilayer transition metal dichalcogenides {MX$_2$ (M= Mo, W; X=
  O, S, Se, Te):} a comparative first-principles study}.
\newblock \emph{\bibinfo{journal}{Chinese Phys. B}}
  \textbf{\bibinfo{volume}{24}}, \bibinfo{pages}{097103}
  (\bibinfo{year}{2015}).

\bibitem{FePS3YoungsHashemi2017}
\bibinfo{author}{Hashemi, A.}, \bibinfo{author}{Komsa, H.-P.},
  \bibinfo{author}{Puska, M.} \& \bibinfo{author}{Krasheninnikov, A.~V.}
\newblock \bibinfo{title}{Vibrational properties of metal phosphorus
  trichalcogenides from first-principles calculations}.
\newblock \emph{\bibinfo{journal}{J. Phys. Chem. C}}
  \textbf{\bibinfo{volume}{121}}, \bibinfo{pages}{27207}
  (\bibinfo{year}{2017}).

\bibitem{ChenThesis2013}
\bibinfo{author}{Chen, C.}
\newblock \emph{\bibinfo{title}{PhD thesis}} (\bibinfo{publisher}{Columbia
  University}, \bibinfo{year}{2013}).

\bibitem{Weber2014}
\bibinfo{author}{Weber, P.}, \bibinfo{author}{G\"{u}ttinger, J.},
  \bibinfo{author}{Tsioutsios, I.}, \bibinfo{author}{Chang, D.~E.} \&
  \bibinfo{author}{Bachtold, A.}
\newblock \bibinfo{title}{Coupling graphene mechanical resonators to
  superconducting microwave cavities}.
\newblock \emph{\bibinfo{journal}{Nano Lett.}} \textbf{\bibinfo{volume}{14}},
  \bibinfo{pages}{2854} (\bibinfo{year}{2014}).

\bibitem{StrainAnisotropyDong2020}
\bibinfo{author}{Dong, X.-J.}, \bibinfo{author}{You, J.-Y.},
  \bibinfo{author}{Zhang, Z.}, \bibinfo{author}{Gu, B.} \& \bibinfo{author}{Su,
  G.}
\newblock \bibinfo{title}{Great enhancement of curie temperature and magnetic
  anisotropy in two-dimensional van der {Waals} magnetic semiconductor
  heterostructures}.
\newblock \emph{\bibinfo{journal}{Phys. Rev. B}}
  \textbf{\bibinfo{volume}{102}}, \bibinfo{pages}{144443}
  (\bibinfo{year}{2020}).

\bibitem{ExchangeBiasZhang2020}
\bibinfo{author}{Zhang, L.} \emph{et~al.}
\newblock \bibinfo{title}{Proximity-coupling-induced significant enhancement of
  coercive field and {C}urie temperature in {2D} van der {Waals}
  heterostructures}.
\newblock \emph{\bibinfo{journal}{Adv. Mater.}} \textbf{\bibinfo{volume}{32}},
  \bibinfo{pages}{2002032} (\bibinfo{year}{2020}).

\bibitem{InterfaceWang2020}
\bibinfo{author}{Wang, Y.} \emph{et~al.}
\newblock \bibinfo{title}{Modulation doping via a two-dimensional atomic
  crystalline acceptor}.
\newblock \emph{\bibinfo{journal}{Nano Lett.}} \textbf{\bibinfo{volume}{20}},
  \bibinfo{pages}{8446--8452} (\bibinfo{year}{2020}).

\bibitem{TauDolleman2017}
\bibinfo{author}{Dolleman, R.~J.} \emph{et~al.}
\newblock \bibinfo{title}{Optomechanics for thermal characterization of
  suspended graphene}.
\newblock \emph{\bibinfo{journal}{Phys. Rev. B}} \textbf{\bibinfo{volume}{96}},
  \bibinfo{pages}{165421} (\bibinfo{year}{2017}).

\bibitem{TauMorell2019}
\bibinfo{author}{Morell, N.} \emph{et~al.}
\newblock \bibinfo{title}{Optomechanical measurement of thermal transport in
  two-dimensional {MoSe$_2$} lattices}.
\newblock \emph{\bibinfo{journal}{Nano Lett.}} \textbf{\bibinfo{volume}{19}},
  \bibinfo{pages}{3143} (\bibinfo{year}{2019}).

\bibitem{MagneticExitationBenShimon2021}
\bibinfo{author}{Ben-Shimon, Y.} \&
  \bibinfo{author}{Ya{\textquotesingle}akobovitz, A.}
\newblock \bibinfo{title}{Magnetic excitation and dissipation of multilayer
  two-dimensional resonators}.
\newblock \emph{\bibinfo{journal}{Appl. Phys. Lett.}}
  \textbf{\bibinfo{volume}{118}}, \bibinfo{pages}{063103}
  (\bibinfo{year}{2021}).

\bibitem{CastellanosGomez2014}
\bibinfo{author}{Castellanos-Gomez, A.} \emph{et~al.}
\newblock \bibinfo{title}{Deterministic transfer of two-dimensional materials
  by all-dry viscoelastic stamping}.
\newblock \emph{\bibinfo{journal}{2D Mater.}} \textbf{\bibinfo{volume}{1}},
  \bibinfo{pages}{011002} (\bibinfo{year}{2014}).

\bibitem{GorbachevWill2017}
\bibinfo{author}{Will, M.} \emph{et~al.}
\newblock \bibinfo{title}{High quality factor graphene-based two-dimensional
  heterostructure mechanical resonator}.
\newblock \emph{\bibinfo{journal}{Nano Lett.}} \textbf{\bibinfo{volume}{17}},
  \bibinfo{pages}{5950} (\bibinfo{year}{2017}).

\end{thebibliography}

\end{document}